\documentclass[mathleft
]{an}
\usepackage{graphicx}
\usepackage{times}
\begin{document}

\Pagespan{789}{}
\Yearpublication{2006}%
\Yearsubmission{2005}%
\Month{11}%
\Volume{999}%
\Issue{88}%

\title{MUCHFUSS - Massive Unseen Companions to Hot Faint Underluminous Stars from SDSS}

\author{S. Geier\inst{1}\fnmsep\thanks{Corresponding author:
  \email{geier@sternwarte.uni-erlangen.de}\newline}
\and  V. Schaffenroth\inst{1}
\and  H. Hirsch\inst{1}
\and  A. Tillich\inst{1}
\and  U. Heber\inst{1}
\and  P. F. L. Maxted\inst{2}
\and  R. H. \O stensen\inst{3}
\and  B. N. Barlow\inst{4,5}
\and  S. J. O'Toole\inst{6}
\and  T. Kupfer\inst{1,7}
\and  T. Marsh\inst{8}
\and  B. G\"ansicke\inst{8}
\and  R. Napiwotzki\inst{9}
\and  O. Cordes\inst{10}
\and  S. M\"uller\inst{1}
\and  L. Classen\inst{1}
\and  E. Ziegerer\inst{1}
\and  H. Drechsel\inst{1}
}
\titlerunning{MUCHFUSS}
\authorrunning{S. Geier et al.}
\institute{Dr. Karl Remeis-Observatory \& ECAP, Astronomical Institute, Friedrich-Alexander University Erlangen-Nuremberg, Sternwartstr. 7, D 96049 Bamberg, Germany
\and 
Astrophysics Group, Keele University, Staffordshire, ST5 5BG, UK
\and 
Institute of Astronomy, K.U.Leuven, Celestijnenlaan 200D, B-3001 Heverlee, Belgium
\and
Department of Physics and Astronomy, University of North Carolina, Chapel Hill, NC 27599-3255, USA
\and 
Department of Astronomy \& Astrophysics, Eberly College of Science, The Pennsylvania State University, 525 Davey Lab, University Park, PA 16802, USA
\and 
Australian Astronomical Observatory, PO Box 296, Epping, NSW, 1710, Australia
\and
Department of Astrophysics, Faculty of Science, Radboud University Nijmegen, P.O. Box 9010, 6500 GL Nijmegen, NE
\and
Department of Physics, University of Warwick, Conventry CV4 7AL, UK
\and
Centre of Astrophysics Research, University of Hertfordshire, College Lane, Hatfield AL10 9AB, UK
\and 
Argelander-Institut f\"ur Astronomie, Auf dem H\"ugel 71, 53121 Bonn, Germany}
\received{}
\accepted{}
\publonline{}


\abstract{%
The project Massive Unseen Companions to Hot Faint Underluminous Stars from SDSS (MUCHFUSS) aims at finding hot subdwarf stars with massive compact companions (white dwarfs with masses $M>1.0\,{\rm M_{\odot}}$, neutron stars or black holes). The existence of such systems is predicted by binary evolution calculations and some candidate systems have been found. We identified $\simeq1100$ hot subdwarf stars from the Sloan Digital Sky Survey (SDSS). Stars with high velocities have been reobserved and individual SDSS spectra have been analysed. About $70$ radial velocity variable subdwarfs have been selected as good candidates for follow-up time resolved spectroscopy to derive orbital parameters and photometric follow-up to search for features like eclipses in the light curves. Up to now we found nine close binary sdBs with short orbital periods ranging from $\simeq0.07\,{\rm d}$ to $1.5\,{\rm d}$. Two of them are eclipsing binaries with companions that are most likely of substellar nature.}

\maketitle

\section{Introduction}

Hot subdwarf stars (sdO/B) are core-helium bur\-ning stars located at the extreme blue end of the horizontal branch (see Heber \cite{heber09} for a review). These stars with masses of about half solar consist almost entirely of helium surrounded by a very thin hydrogen envelope only. The cause of the extreme mass-loss in the red-giant phase necessary to form hot subdwarfs remains unclear. 

A significant fraction ($\simeq50\%$) of the sdBs stars are in short period binaries (Maxted et al. \cite{maxted01}; Napiwotzki et al. \cite{napiwotzki04}) with periods ranging from only $0.07\,{\rm d}$ to more than $10\,{\rm d}$. These close binary sdBs are most likely formed by common envelope (CE) ejection (Han et al. \cite{han02,han03}). Because most of them are single-lined, only lower mass limits have been derived from the binary mass functions consistent with late main sequence stars of spectral type M or compact objects like white dwarfs (WDs). 

Subdwarf binaries with massive WD companions may be candidates for supernova type Ia (SN~Ia) progenitors because these systems lose angular momentum due to the emission of gravitational waves and shrink. Mass transfer or the subsequent merger of the system may cause the WD to reach the Chandrasekhar limit and explode as a SN~Ia. One of the best known candidate systems for the double degenerate merger scenario is the sdB+WD binary KPD\,1930$+$2752 (Maxted et al. \cite{maxted00}; Geier et al. \cite{geier07}). 

Geier et al. (\cite{geier10a,geier10b}) analysed high resolution spectra of single-lined sdB binaries. Because the inclinations of these systems are not known, additional information is needed to derive companion masses. Accordingly, Geier et al. (\cite{geier10a,geier10b}) performed a quantitative spectral analysis and determined surface gravities and projected rotational velocities. Assuming synchronised orbits the masses and the nature of the unseen companions were constrained. Surprisingly, the masses of some companions are only consistent with either massive white dwarfs ($M>1.0\,{\rm M_{\odot}}$), neutron stars (NS) or stellar mass black holes (BH). However, the assumption of orbital synchronisation in close sdB binaries was shown to be not always justified and the analysis suffers from selection effects (Geier et al. \cite{geier10b}). On the other hand, the existence of sdB+NS/BH systems is predicted by binary evolution theory (Podsiadlowski et al. \cite{podsi02}; Pfahl et al. \cite{pfahl03}; Yungelson \& Tutukov \cite{yungelson05}; Nelemans \cite{nelemans10}). The formation channel includes two phases of unstable mass transfer and one supernova explosion, while  the fraction of systems formed in this way is predicted to be about $1-2\%$ of all sdBs. 

If the companion were a neutron star, it could be detectable by radio observations as a pulsar. Coenen et al. (\cite{coenen11}) searched for pulsed radio emission at the positions of four candidate systems from Geier et al. (\cite{geier10b}) using the Green Bank radio telescope, but did not detect any signals. Most recently, Mereghetti et al. (\cite{mereghetti11}) searched for X-ray signatures of mass transfer driven by weak stellar winds from the sdBs. Using the XRT instrument on board of the SWIFT satellite and targeting twelve binaries from the sample of Geier et al. (\cite{geier10b}), Mereghetti et al. (\cite{mereghetti11}) did not detect any X-ray emission.

In order to find sdBs with compact companions like massive white dwarfs ($M>1.0\,M_{\rm \odot}$), neutron stars or black holes we started a radial velocity (RV) survey (Massive Unseen Companions to Hot Faint Underluminous Stars from SDSS\footnote{Sloan Digital Sky Survey}, MUCHFUSS, Geier et al. \cite{geier11a,geier11b}). 

The same selection criteria that we applied to find such binaries are also well suited to single out hot sub\-dwarf stars with constant high radial velocities in the Galactic halo like extreme population II and hypervelocity stars and led to a very interested spin-off project (Tillich et al. \cite{tillich11}).

\section{Target selection}

\subsection{Colour and RV selection}

The target selection is optimised to find close massive compact companions to sdB stars. The SDSS spectroscopic database (Data Release 6) is the starting point for our survey.  SdO/B candidates were selected by applying a colour cut to SDSS photometry. All point source spectra within the colours $u-g<0.4$ and $g-r<0.1$ were downloaded from the SDSS Data Archive Server\footnote{das.sdss.org}. About $10\,000$ hot stars were classified by visual inspection. The sample contains $1100$ hot subdwarfs (for details see Geier et al. \cite{geier11a}). 

SdBs with radial velocities (RVs) lower than $\pm100\,{\rm km\,s^{-1}}$ have been excluded to filter out such binaries with normal disc kinematics, by far the majority of the sample. Another selection criterion is the brightness of the stars since the quality of the spectra is not sufficient for faint stars. Because of that most objects much fainter than $g=19\,{\rm mag}$ have been excluded. 

\subsection{Survey for RV variable stars}

Second epoch medium resolution spectroscopy ($R=1800-4000$) was obtained for $88$ stars using ESO-VLT/FORS1, WHT/ISIS, CAHA-3.5m/TWIN and ESO-NTT/EFOSC2. Second epoch observations by SDSS have been used as well. We discovered $46$ RV variable systems in this way. 

\begin{figure*}
\begin{center}
  \includegraphics[width=13cm]{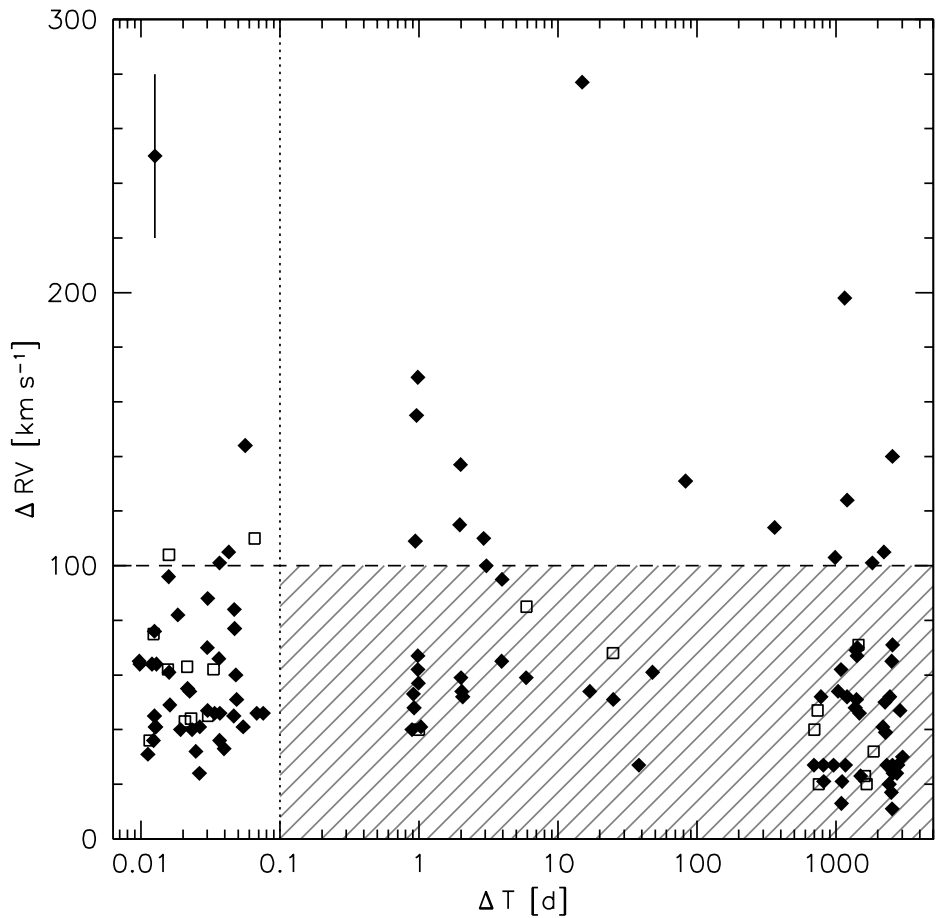}
\end{center}
\caption{Highest radial velocity shift between individual spectra ($\Delta RV$) plotted against time difference between the corresponding observing epochs ($\Delta T$). The dashed horizontal line marks the selection criterion $\Delta RV>100\,{\rm km\,s^{-1}}$, the dotted vertical line the selection criterion $\Delta T<0.1\,{\rm d}$. All objects fulfilling at least one of these criteria lie outside the shaded area and belong to the top candidate list for the follow-up campaign. The filled diamonds mark sdBs, while the open squares mark He-sdOs (Geier et al. (\cite{geier11a}).}
\label{fig:dTdRV}
\end{figure*}

The SDSS spectra are co-added from at least three individual ``sub-spectra'' with typical exposure times of $15\,{\rm min}$, which are normally taken consecutively. Hence, those spectra can be used to probe for radial velocity variations on short timescales. We have obtained the sub-spectra for all sdBs brighter than $g=18.5\,{\rm mag}$ and discovered $81$ new sdB binaries with radial velocity variations on short time scales ($\simeq0.03\,{\rm d}$) in this way. In total we found $127$ new RV variable hot subdwarf stars (see Fig.~\ref{fig:dTdRV}).

In addition, $20$ helium-rich sdOs (He-sdOs) show RV variability. This fraction was unexpectly high since in the SPY sample only $4\%$ of these stars turned out to be RV variable (Napiwotzki et al. \cite{napiwotzki08}). However, it is not yet clear what causes this RV variability, since the orbital parameters of any such object couldn't be derived yet. This would be necessary to prove, that  these He-sdOs are in close binaries. 

\subsection{Selection of candidates with massive companions}

Numerical simulations were carried out to select the most promising targets for follow-up and estimate the probability for a subdwarf binary with known RV shift to host a massive compact companion. We created a mock sample of sdBs with a close binary fraction of $50\,\%$ and adopted the distribution of orbital periods of the known sdB binaries. Two RVs were taken from the model RV curves at random times and the RV difference was calculated for each of the $10^{6}$ binaries in the simulation sample. To account for the fact that the individual SDSS spectra were taken within short timespans, another simulation was carried out, where the first RV was taken at a random time, but the second one just $0.03\,{\rm d}$ later (for details see Geier et al. \cite{geier11a}). 

The sample of promising targets consists of $69$ objects in total. These objects either show significant RV shifts ($>30\,{\rm km\,s^{-1}}$) within $0.03\,{\rm d}$ ($52$ stars) or high RV shifts ($100-300\,{\rm km\,s^{-1}}$) within more than one day ($17$ stars). An extension of our target selection to SDSS Data Release 7 is in progress.

\section{Sample statistics}

The classification of the hot subdwarf sample is based on existence, width, and depth of helium and hydrogen absorption lines as well as the flux distribution between $4000$ and $6000\,{\rm \AA}$. Subdwarf B stars show broadened hydrogen Balmer and He\,{\sc i} lines, sdOB stars He\,{\sc ii} lines in addition, while the spectra of sdO stars are dominated by weak Balmer and strong He\,{\sc ii} lines depending on the He abundance. A flux excess in the red compared to the reference spectrum as well as the presence of spectral features such as the Mg\,{\sc i} triplet at $5170\,{\rm \AA}$ or the Ca\,{\sc ii} triplet at $8650\,{\rm \AA}$ are taken as indications of a late type companion.

In total we found $1100$ hot subdwarfs. $725$ belong to the class of single-lined sdBs and sdOBs. Features indicative of a cool companion were found for $89$ of the sdBs and sdOBs. $9$ sdOs have main sequence companions, while $198$ sdOs, most of which show helium enrichment, are single-lined (Geier et al. \cite{geier11a}).

\section{Spectroscopy follow-up}

Follow-up medium resolution ($R=1200-4000$) spectra were taken during dedicated follow-up runs with ESO-NTT/EFOSC2, WHT/ISIS, CAHA-3.5m/TWIN, INT/IDS, SOAR/Goodman and Gemini-N/GMOS. Orbital parameters of eight sdB binaries discovered in the course of the MUCHFUSS project have been determined so far Geier et al. (\cite{geier11b,geier11c}). 

Because we deal with single-lined spectroscopic binaries, only their mass functions $f_{\rm m} = M_{\rm comp}^3 \sin^3i/(M_{\rm comp} + M_{\rm sdB})^2 = P K^3/2 \pi G$ can be calculated. Although the RV semi-amplitude $K$ and the period $P$ can be derived from the RV curve, the sdB mass $M_{\rm sdB}$, the companion mass $M_{\rm comp}$ and the inclination angle $i$ remain free parameters. Adopting the canonical mass for core helium-burning stars $M_{\rm sdB}=0.47\,M_{{\rm \odot}}$ and $i<90^{\rm \circ}$ we derive a lower limit for the companion mass. 

With this minimum mass a qualitative classification of the companions' nature is possible in certain cases. For mini\-mum companion masses lower than $0.45\,M_{\rm \odot}$ a main sequence companion can not be excluded because its luminosity would be too low to be detectable in the optical spectra (Lisker et al. \cite{lisker05}). The companion could therefore be a compact object like a WD or a late main sequence star. If the minimum companion mass exceeds $0.45\,M_{\rm \odot}$ and no spectral signatures of the companion are visible, it has to be a compact object. If the mass limit exceeds $1.00\,M_{\rm \odot}$ or even the Chandrasekhar limit ($1.40\,M_{\rm \odot}$) the existence of a massive WD or even an NS or BH companion is proven.

The derived minimum companion masses of seven binaries from our sample are similar ($0.32-0.41\,M_{\rm \odot}$). From these minimum masses alone the nature of the companions cannot be constrained unambiguously. However, the fact that all seven objects belong to the sdB binary population with the highest minimum masses illustrates that our target selection is efficient and singles out sdB binaries with massive companions (see Geier et al. \cite{geier11b}). 

\section{Photometry follow-up}

Photometric follow-up helps to clarify the nature of the companions. Short period sdB binaries with late main sequence or substellar companions show variations in their light curves caused by the irradiated surfaces of the cool companions facing the hot subdwarf stars. If this so-called reflection effect is present, the companion is most likely a main sequence star. If not, the companion is most likely a compact object. In the case of the short period system J1138$-$0035 a light curve taken by the SuperWASP project (Pollacco et al. \cite{pollacco06}) shows no variation exceeding $\simeq1\%$. The companion is therefore most likely a white dwarf (Geier et al. \cite{geier11b}).

\begin{figure*}
\begin{center}
  \includegraphics[width=10cm]{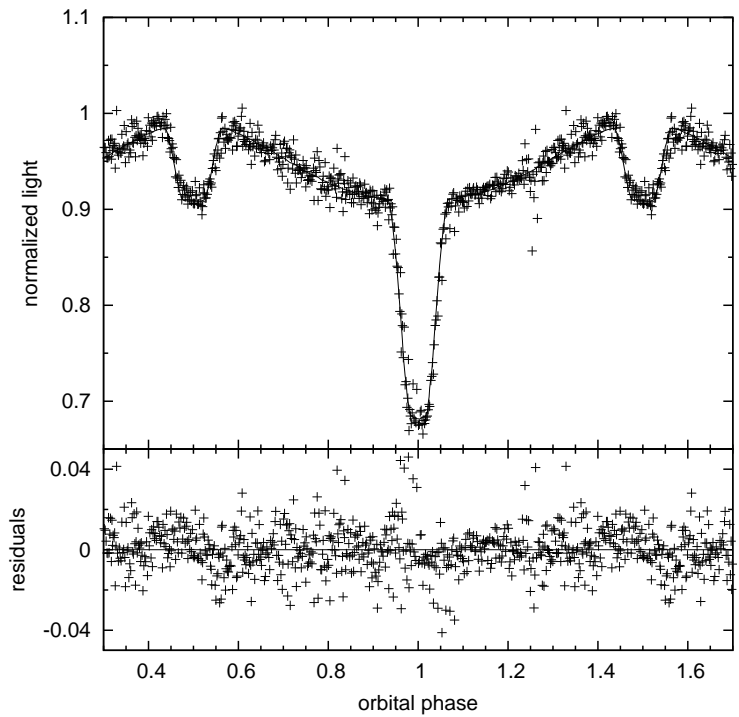}
\end{center}
\caption{Phased R-band light curve of J0820+0008 taken with the Mercator telescope. A light curve model is fitted to the data and residuals are given below. Primary and secondary eclipses can be clearly seen as well as the sinusoidal shape caused by the reflection effect (Geier et al. \cite{geier11c}).}
\label{fig:J0820}
\end{figure*}

We obtained follow-up photometry with the Mercator telescope and the BUSCA instrument mounted on the CAHA-2.2m telescope. In this way we discovered the first eclipsing sdB binary J0820+0008 to host a brown dwarf companion with a mass ranging from $0.045$ to $0.068\,M_{\rm \odot}$ (Fig.~\ref{fig:J0820}, Geier et al. \cite{geier11c}). 

\begin{figure*}
\begin{center}
  \includegraphics[width=9cm, angle=-90]{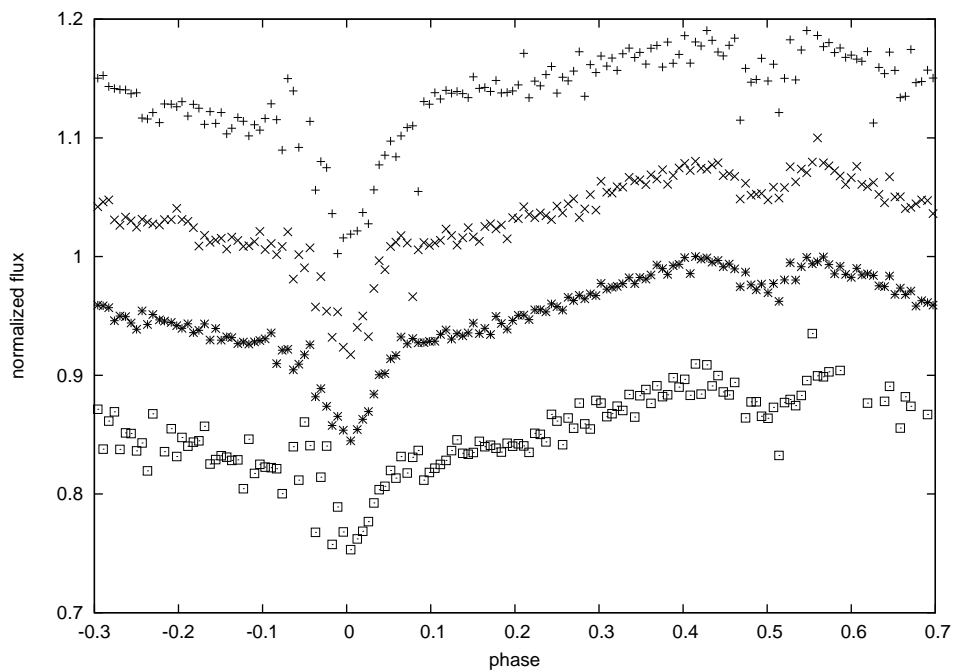}
\end{center}
\caption{Phased light curves of J1622+4730 taken with BUSCA (UV,B,R,IR-band). Although the eclipses are not total, the light curve looks very similar to the one of J0820+0008 (see Fig.~\ref{fig:J1622}).}
\label{fig:J1622}
\end{figure*}

A very similar eclipsing system (J1622+4730) was discovered serendipituously (see Fig.~\ref{fig:J1622}). A preliminary analysis shows that the orbital period is very short ($\simeq0.07\,{\rm d}$) and the RV semi-amplitude quite low ($\simeq47\,{\rm km\,s^{-1}}$). The companion is most likely a substellar object as well. The high success rate in finding these objects shows that our target selection not only singles out sdB binaries with high RV-amplitudes, but also systems with very short orbital periods and moderate RV-amplitudes. 

Most recently, we detected p-mode pulsations in the sdB J0120+3950 (FBS\,0117+396, Geier et al. \cite{geier11a}) as well as a longer trend indicative of a reflection effect in a light curve taken with BUSCA. Only a few of the known short-period sdB pulsators (sdBV$_{\rm r}$) are in close binary systems. 

\section{Substellar companions}

The existence of eclipsing sdB+ late dM binaries of HW\,Vir type with short orbital periods ($0.1-0.26\,{\rm d}$) and companion masses between $0.1\,M_{\rm \odot}$ and $0.2\,M_{\rm \odot}$ (For et al. \cite{for10}; \O stensen et al. \cite{oestensen10} and references therein) shows that stars close to the nuclear burning limit of $\simeq0.08\,M_{\rm \odot}$ are able to help eject a common envelope and form a hot subdwarf. Substellar companions to sdB stars have been found using the light travel time technique (Schuh \cite{schuh10} and references therein). However, these systems have wide orbits and none of these companions influenced the evolution of its host star. 

In the course of the MUCHFUSS project we discovered two sdBs with most likely substellar companions in close orbits. These companions evidently interacted with the sdB progenitor stars and caused the ejection of the common envelopes. Some theoretical models indeed predict such an interaction between planets or brown dwarf companions and their nearby host stars to be a possible formation channel for hot subdwarfs and helium white dwarfs (Soker \cite{soker98}; Nelemans \& Tauris \cite{nelemans98}). Our finding can be used to constrain such models and learn more about the role substellar companions play in the formation of single and close binary sdBs.

\section{Summary}

The MUCHFUSS project aims at finding hot subdwarf stars with massive compact companions. We identified 1100 hot subdwarfs by colour selection and visual inspection of the SDSS-DR6 spectra. The best candidates for massive compact companions are followed up with time resolved medium resolution spectroscopy. Up to now orbital solutions have been found for eight single-lined binaries. Seven of them have large minimum companion masses compared to the sample of known close binaries, which shows that our target selection works quite well. However, it turns out that our selection strategy also allows us to detect low-mass companions to sdBs in very close orbits. We discovered an eclipsing sdB with a brown dwarf companion and a very similar candidate system in the course of our photometric follow-up campaign. These early results encourage us to go on, because they demonstrate that MUCHFUSS will find both the most massive and the least massive companions to sdB stars.

\end{document}